\documentclass{aa}  

\usepackage{natbib,twoopt}
\usepackage[breaklinks=true]{hyperref} 
\bibpunct{(}{)}{;}{a}{}{,}    
\hypersetup{colorlinks=true, linkcolor=blue, filecolor=magenta, urlcolor=blue, citecolor = blue }

\usepackage{graphicx}
\usepackage{txfonts}
\begin{document} 

\titlerunning{Whistler instability driven by the sunward electron deficit}

\title{Whistler instability driven by the sunward electron deficit in the solar wind}

\subtitle{High-cadence Solar Orbiter observations}

\author{L. Ber\v{c}i\v{c} \inst{1}
\and D. Verscharen \inst{1,2}
\and C. J. Owen \inst{1}
\and L. Colomban \inst{3}
\and M. Kretzschmar \inst{3}
\and T. Chust \inst{4}
\and M. Maksimovic \inst{5}
\and D.~O. Kataria \inst{1}
\and C. Anekallu \inst{1}
\and E. Behar \inst{6,7}
\and M. Berthomier \inst{4}
\and R. Bruno \inst{8}
\and V. Fortunato \inst{9}
\and C. W. Kelly \inst{1}
\and Y.~V. Khotyaintsev \inst{10}
\and G.~R. Lewis \inst{1}
\and S. Livi \inst{11}
\and P. Louarn \inst{12}
\and G. Mele \inst{13}
\and G. Nicolaou \inst{11}
\and G. Watson \inst{1}
\and R. T. Wicks \inst{14}
}

\institute{Mullard Space Science Laboratory, University College London, Dorking, RH5 6NT, UK
\and Space Science Center, University of New Hampshire, 8 College Road, Durham NH 03824, USA
\and LPC2E/CNRS, 3 Avenue de la Recherche Scientifique, 45071 Orléans Cedex 2, France
\and LPP, CNRS, Ecole Polytechnique, Sorbonne Universit\'{e}, Observatoire de Paris, Universit\'{e} Paris-Saclay, Palaiseau, Paris, France
\and LESIA, Observatoire de Paris, Université PSL, CNRS, Sorbonne Université, Univ. Paris Diderot, Sorbonne Paris Cité, 5 place Jules Janssen, 92195 Meudon, France
\and Swedish Institute of Space Physics (IRF), Kiruna, Sweden
\and Laboratoire Lagrange, OCA, UCA, CNRS, Nice, France
\and INAF-IAPS, Via Fosso del Cavaliere 100, 00133 Roma, Italy
\and Planetek Italia, Italy
\and Swedish Institute of Space Physics (IRF), Uppsala, Sweden
\and Southwest Research Institute, San Antonio, Texas, USA
\and IRAP, Université de Toulouse, CNRS, UPS, CNES, Toulouse, France
\and Leonardo, Taranto, Italy
\and Department of Mathematics, Physics and Electrical Engineering, Northumbria University, Newcastle upon Tyne, UK
}
   
\date{Received March 31, 2021; accepted April 15, 2021}

 
  \abstract
   {Solar wind electrons play an important role in the energy balance of the solar wind acceleration by carrying energy into interplanetary space in the form of electron heat flux. The heat flux is stored in the complex electron velocity distribution functions (VDFs) shaped by expansion, Coulomb collisions, and  field-particle interactions.}
   {We investigate how the suprathermal electron deficit in the anti-strahl direction, which was recently discovered in the near-Sun solar wind, drives a kinetic instability and creates whistler waves with wave vectors that are quasi-parallel to the direction of the background magnetic field.}
   {We combine high-cadence measurements of electron pitch-angle distribution functions and electromagnetic waves provided by Solar Orbiter during its first orbit. Our case study is based on a burst-mode data interval from the Electrostatic Analyser System (SWA-EAS) at a distance of  112 $R_S$ (0.52 au) from the Sun, during which several whistler wave packets were detected by Solar Orbiter's Radio and Plasma Waves (RPW) instrument. }
   {The sunward deficit creates kinetic conditions under which the quasi-parallel whistler wave is driven unstable. We directly test our  predictions for the existence of these waves through solar wind observations. We find whistler waves that are quasi-parallel and almost circularly polarised, propagating away from the Sun, coinciding with a pronounced sunward deficit in the electron VDF. The cyclotron-resonance condition is fulfilled for electrons moving in the direction opposite to the direction of wave propagation, with energies corresponding to those associated with the sunward deficit. }
   {We conclude that the sunward deficit acts as a source of quasi-parallel whistler waves in the solar wind. The quasilinear diffusion of the resonant electrons tends to fill the deficit, leading to a reduction in the total electron heat flux.}


 \maketitle
%

\section{Introduction}

The thermal energy of the solar corona is sufficient to accelerate part of its plasma into interplanetary space and create the solar wind. The solar wind consists mainly of protons, which carry the solar wind mass and momentum flux, and electrons, which carry the solar wind heat flux.

The heat flux is measured as the skewness (i.e., the third moment) of the electron velocity distribution function (VDF), which evolves with radial distance. The electron VDF typically consists of three separate populations. Most of the electrons (more than 90\%) belong to the core population which is present at low electron energies and dominated by Coulomb collisions. Two suprathermal populations exist at higher energies: the strahl, which represents electrons typically streaming away from the Sun along the magnetic field, and the halo present at all pitch angles. The relative drifts of these populations must ensure the zero-current condition in the proton-rest frame: $n_cv_c + n_sv_s + n_hv_h = 0$, where $n_j$ are the densities and $v_j$ are the bulk velocities of the core ($j=c$), strahl ($j=s$), and halo ($j=h$) \citep{Feldman1975a, SchwartcMarsch1983, Pilipp1987b, Maksimovic1997b, Maksimovic2005a, stverak2008, Stverak2009, Tao2016}. 

Different models to capture the properties of the three electron populations have been proposed in the past, often chosen based on the capability of the data set used, and the needs in term of research goals \citep[e.g.][]{Stverak2009, Horaites2018a, Bercic2020}. The core electrons are often represented by a bi-Maxwellian distribution oriented with respect to the magnetic field. Other models include, for example, bi-$\kappa$ distributions and bi-self-similar distributions, which better describe the core electrons observed by the Wind spacecraft in the vicinity of interplanetary shocks at 1 au \citep{Wilson2019a, Wilson2019b}.

Recent observations of the near-Sun solar wind by Parker Solar Probe reveal a departure from the Maxwellian fit for the core population in the shape of a sunward deficit. This deficit is aligned with the magnetic field and located in the direction opposite to the strahl / heat flux in velocity space \citep{Halekas2019, Halekas2020, Halekas2021, Bercic2020, Bercic2021ambip}. \citet{Halekas2020} show that the deficit makes a significant contribution to the total electron heat flux.

Exospheric models of the solar wind describe the acceleration of the solar wind as the result of the ambipolar diffusion of protons and electrons in an atmosphere that transitions from collisional to collisionless conditions at the exobase. The electrons are, due to their smaller mass, much more mobile than the protons. This creates an ambipolar electric field which assures the equality of electron and proton fluxes. In exospheric models, the electron VDF consists of a highly anisotropic core and a very narrow strahl, separated in energy by the ambipolar potential energy. In the sunward direction, the electron core exhibits an abrupt cutoff defined by the ambipolar potential \citep{Lemaire1970, LemaireJosephandScherer1971, Jockers1970, Maksimovic1997c, Pierrard1999, Zouganelis2004a}.

Observations differ from these predictions of collisionless exospheric models. For example, solar wind core electrons are often quasi-isotropic, the strahl is substantially scattered towards larger pitch angles, and a second suprathermal population, the halo, is present. The differences between the predicted and observed VDFs are attributed to the effects of Coulomb collisions and field-particle interactions. Coulomb collisions are efficient in isotropising the core population \citep{LieSvendsen1997, pierrard2001, Salem2003, Smith2012, stverak2008} as well as scattering the strahl at low electron energies \citep{Horaites2018a, Horaites2019, BoldyrevHoraites2019, Bercic2021}. At higher energies, where collisions are rare, kinetic instabilities can reduce the skewness of electron VDF \citep{Hollweg1974, Gary1975b, Feldman1976, Lakhina1977, Krafft2005, SaitoGary2007}.

Resonant, electron-driven instabilities typically create waves with frequencies between the ion and electron gyrofrequency. Enhanced fluctuations in this frequency band in the solar wind often correspond to whistler waves. The most common observed type of whistler waves are right-hand polarised, quasi-parallel whistler waves \citep{Lacombe2014, Tong2019stat, Jagarlamudi2020, Jagarlamudi2021}. \citet{Zhang1998, Stansby2016} and \citet{Kretzschmar2021} further find that the majority of the detected waves propagate in the anti-sunward direction. These waves only resonate with electrons through the cyclotron resonance, which is fulfilled by electrons moving in the opposite direction of the wave phase speed along the magnetic field. Whistler waves of this type cannot contribute to the scattering of the strahl electrons \citep{Verscharen2019scat}. 

Less frequently observed sunward whistler waves may be present in the solar wind, but obscured by turbulent fluctuations \citep{Vasko2020, Vocks2012}. Oblique whistler waves, are also able to interact with strahl electrons and scatter them through the strahl-driven oblique whistler instability \citep{Vasko2019, Verscharen2019scat, Agapitov2020, Micera2020}. Oblique whistlers were recently observed in the near-Sun solar wind \citep{Agapitov2020,Cattell2020} and associated with enhanced strahl pitch-angle widths \citep{Cattell2021}.

We propose a resonant instability mechanism that creates quasi-parallel whistler waves through the relaxation of the sunward deficit in the core energy range in section \ref{sect:prediction}. In Section \ref{sec:data} we describe the analysis of the high-cadence electron VDFs and electromagnetic waves measured by Solar Orbiter. In Section \ref{sec:observations} we display the observational results and relate them to the proposed instability mechanism. In section \ref{sec:discussion} we discuss how our prediction corresponds to the data and existing studies and summarise our findings.

\section{The quasi-parallel whistler instability driven by the sunward deficit}
\label{sect:prediction}

Recent in situ solar wind observations indicate that core electrons with energies larger than $\sim$ 2 $w_{\mathrm c\parallel}$ depart from a traditional bi-Maxwellian model, and exhibit a deficit in the direction opposite to the heat flux \citep{Halekas2019, Halekas2020, Halekas2021, Bercic2020, Bercic2021ambip}. We investigate whether the sunward deficit in the electron VDF can drive whistler waves unstable, and release its free energy through the growth of electromangetic fluctuations.

One mechanism that allows an energy exchange between particles and fields are the resonant interactions between particles and waves. These interactions are characterised by a resonance condition. In the case of parallel-propagating whistler waves with frequency $\omega$ and wavenumber $k_{\parallel}$, the resonance condition is \citep[e.g.][]{SagdeevGaleev1969, Marsch2006,ShklyarMatsumoto2009}
\begin{equation}
v_{\parallel}=v_{\mathrm{cyclo}}\equiv \frac{\omega-\omega_\mathrm{ ce}}{k_{\parallel}},
    \label{eq:res}
\end{equation}
where $\omega_\mathrm{ce}=eB/m_\mathrm e$ is the electron gyrofrequency, $B$ is the magnetic field, and $m_\mathrm e$ is the electron mass.
Only electrons with the parallel velocity component $v_{\parallel}=v_{\mathrm{cyclo}}$ undergo resonant interactions. Landau-resonant or higher-order cyclotron-resonant interactions are not accessible for parallel-propagating whistler waves \citep[e.g.][]{SagdeevGaleev1969, Verscharen2019_review}.

Quasi-parallel whistler waves are circular, purely right-hand polarised waves with frequencies $\omega < \omega_\mathrm{ce}$. Therefore, $v_{\mathrm{cyclo}}<0$ if $\omega/k_{\parallel}>0$ and vice versa according to Eq.~(\ref{eq:res}). This means that the resonant electrons must move in the direction opposite to the wave phase speed.

If electrons fulfil the resonance condition in Eq.~(\ref{eq:res}), they follow diffusion paths according to quasilinear theory. These diffusion paths correspond to curves of constant kinetic energy in the wave rest frame. The ambiguity in the direction of the quasilinear diffusion is resolved by the requirement that the diffusion is always directed from higher phase-space density to lower phase-space density \citep{ShapiroShevchenko1963, KennelEngelmann1966, Marsch2006, ShklyarMatsumoto2009,Verscharen2019scat}. 

The instability mechanism is the same as for the Whistler Heat Flux Instability (WHFI) proposed by \citet{Gary1975b}: the key ingredient are the gradients of the distribution function in velocity space. However, in the classical derivation of the WHFI, these gradients are produced by the relative drift between two electron populations (e.g. the core and the halo). In our case, the pitch angle gradient ($\partial f / \partial \alpha$) is a result of the local shape of the distribution at the resonant speed.

\begin{figure}
    \includegraphics[width=0.5\textwidth]{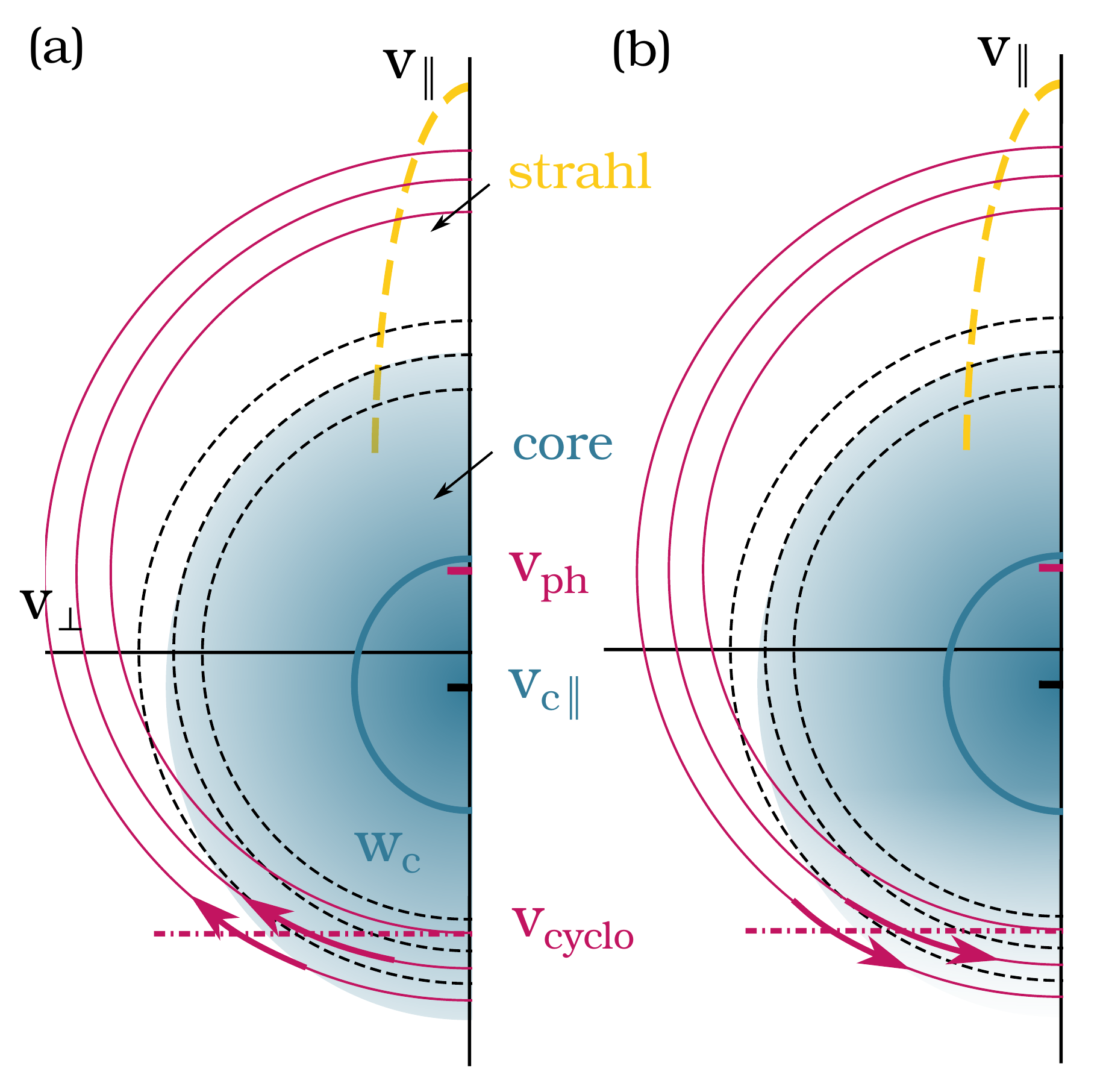}
    \caption{A schematics of the whistler instability driven by the sunward deficit in the anti-strahl direction. The core population is shown as a blue shaded semi-circle, with a blue semi-circle marking its thermal velocity ($w_c$) and $v_{c\parallel}$ its bulk velocity in plasma rest frame. The dashed yellow curve indicates the strahl. Pink semi-circles mark curves of constant energy in the rest frame of the wave moving with $v_{ph}=\omega/k_{\parallel}$, and black dashed semi-circles the curves of constant kinetic energy in the plasma frame. Pink arrows indicate the electron diffusion paths.
    (a) Example without the sunward deficit; (b) The sunward deficit can trigger an instability. }
    \label{fig:schematics}
\end{figure}

The schematic in Fig.~\ref{fig:schematics} illustrates  the quasilinear diffusion of electrons in the energy range corresponding to the observed sunward deficit. We show two cases: (a) where electron core is represented by a bi-Maxwellian distribution, and (b) where this distribution exhibits a deficit at $v_{\parallel}<0$. 

The whistler wave assumed in this illustration propagates with $v_\mathrm{ph}=\omega/k_{\parallel}>0$ in the direction of the strahl electrons. It can resonate with the anti-strahl directed electrons with velocity close to $v_\mathrm{cyclo}$ according to Eq.~(\ref{eq:res}) and marked in the figure. The curves of constant energy in the wave frame are shown in pink, and the curves of constant energy in the plasma frame in black. 

The pink arrows describe the direction of the diffusion paths. These diffusion paths are always directed along the pink semi-circles of constant energy in the wave frame for the given $v_\mathrm{ph}$ of the resonant waves. At the same time, the diffusion paths point into the direction opposite to the pitch-angle gradient ($-\partial f / \partial \alpha$). In panel (a) of Fig.~\ref{fig:schematics}, this setup leads to a diffusion of resonant electrons from smaller to larger $v_\perp$. Electrons following these diffusion paths gain kinetic energy in the plasma frame, which must be provided from the assumed whistler wave. The wave amplitude thus decreases, corresponding to wave damping. A resonant wave--particle interaction of this type in which  energy is transferred from the waves to the electrons was studied with a test-particle approach by \citet{Behar2020}. This work finds  that a deficit is created in the electron VDF near $v_\parallel \lesssim v_{\mathrm{cyclo}}$ \citep[see Fig.~4 by][]{Behar2020}. The electrons migrating from this phase-space region diffuse along the semi-circle and create an overdensity at larger pitch-angles with $v_\parallel\gtrsim v_{\mathrm{cyclo}}$.

Fig.~\ref{fig:schematics} (b) shows the case in which the sunward deficit at $v_{\parallel}<0$ creates conditions for the diffusion towards smaller $v_{\perp}$, due to the change in the sign of $\partial f/\partial \alpha$ at $v_{\parallel}=v_{\mathrm{cyclo}}$.  Electrons following the indicated diffusion paths lose kinetic energy. This kinetic energy is transferred into the resonant whistler waves, and causes the waves to grow. Case (b) thus illustrates the instability mechanism of the quasi-parallel whistler instability driven by the sunward electron deficit in the anti-strahl direction.

We predict that this resonant wave--particle mechanism occurs in the solar wind. If this prediction is valid, we anticipate a correlation between the presence of quasi-parallel whistler waves and the presence of increased sunward deficits in the electron VDFs. Our aim is to test this prediction with data from Solar Orbiter.

\section{Data Analysis Methods}
\label{sec:data}
\subsection{Solar wind electrons}

We present in-situ measurements from the first cruise-phase orbit of Solar Orbiter (SO), the latest heliospheric mission designed to link the solar wind to the plasma conditions at its origin in the solar corona \citep{Muller2020}. Solar-wind electrons are measured by the Electrostatic Analyser System (EAS) on board SO, consisting of two top-hat analyser heads, EAS 1 and EAS 2. EAS is part of the Solar Wind Analyser (SWA) instrument suite, which also includes the Proton-Alpha Sensor (PAS), and the Heavy Ion Sensor (HIS) characterising the solar-wind ion populations \citep{Owen2020}. SO's fluxgate magnetometer (MAG) measures the magnetic field with a cadence of 8 Hz \citep{Horbury2020}, while a search-coil magnetometer and electric field antennas, belonging to the Radio and Plasma Waves (RPW) instrument \citep{Maksimovic2020}, cover the higher-frequency magnetic and electric field fluctuations.

\begin{figure}
    \includegraphics[width=0.5\textwidth]{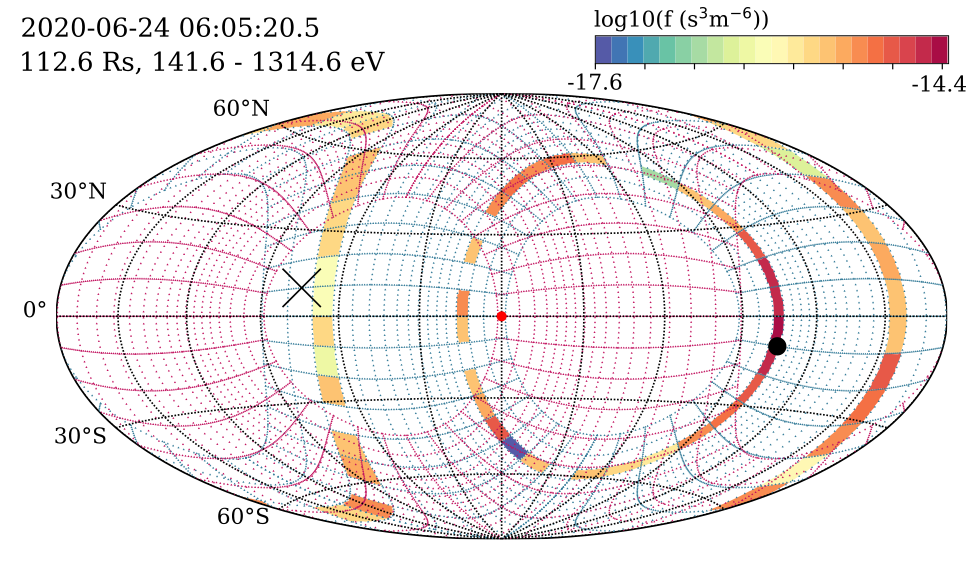}
    \caption{A skymap representation of the combined EAS FOV in the spacecraft (SC) reference frame. The centre of the plot is aligned with the X-axis, 90$^\circ$ in longitude with the Y-axis, and latitude corresponds to the Z-axis. The angular bins are shown in blue for EAS 1 and in red for EAS 2. Two elevation bins of the EAS 1 head are filled with data obtained during one scan in the instrument Burst Mode (BM). The magnetic field direction is marked with a black dot ($+\vec B$) and a black cross ($-\vec B$). }
    \label{fig:fov}
\end{figure}

Each of EAS's instrument heads measures electron inflow directions through 32 azimuth anodes with an angular width of $11.25^\circ$ and 16 elevation deflector states with slightly variable angular widths of $\sim 3 - 10^\circ$. The EAS heads are positioned at the end of the spacecraft main boom, forming a combined field of view (FOV) which covers almost the full solid angle of 4$\pi$. The instrument's FOV is presented in Fig.~\ref{fig:fov} with a skymap plot in the spacecraft (SC) reference frame. In this frame, the X-axis is the longitudinal axis of SO, pointing in the sunward direction, the Y-axis is the transverse azimuthal SC axis, and the Z-axis is orthogonal to the two axes and pointing northward. The blue grid describes the angular bins of EAS 1, and the red grid describes the angular bins of EAS 2. While some parts of the sky are covered by only one of the two heads, the FOVs of EAS 1 and EAS 2 overlap in a large region. 

One electron energy sweep is conducted in 64 exponentially spaced steps, detecting electrons with energies up to 5\,keV, with a relative energy resolution of $\Delta E/E = 0.135$. A full 3D distribution scan is obtained within 0.92\,ms, however downlinked with a much lower cadence due to the limited available telemetry budget (in Normal Mode (NM) every 10\,s or 100\,s). 

In this work, we present electron VDFs  measured in the instrument's Burst Mode (BM) at a cadence of 0.125\,s. This higher time resolution is made possible by a new operational concept, applied for the first time on SO \citep{Owen2020, Owen2021}. Assuming that the measured electron VDFs are gyrotropic, a 3D VDF can be fully described by a 2D VDF in the magnetic field aligned frame. Removing one dimension substantially reduces the VDF data volume and thus the measurement time, as sampling of all perpendicular directions to the magnetic field is omitted. Therefore, before the start of each measurement sequence, SWA receives the information about the current magnetic field vector from MAG. The instrument then defines the EAS head as well as the appropriate elevation deflection states that sample the positive and negative magnetic field directions. The energy sweep is then only performed in the two selected deflection states, for which EAS obtains all azimuth directions simultaneously.
This procedure repeats every 0.125\,s, producing the second-fastest electron pitch-angle distribution measurements on any space mission to date, after the Fast Plasma Investigation (FPI) instrument on-board the Magnetospheric Multiscale (MMS) spacecraft \citep{Burch2016}. A detailed description of the instrument design and operating modes is given by \citet{Owen2020}. The BM operation and first results are presented by \citet{Owen2021}.

We show an example of a BM scan in Fig.~\ref{fig:fov}. We indicate the positive magnetic field direction with a black dot, and the negative magnetic field direction with a black cross. For this case, the data are collected by EAS 1 with the 3rd and the 14th elevation deflection states. Around the positive magnetic field direction, the increased flux represents strahl electrons.

\begin{figure*}
    \includegraphics[width=1\textwidth]{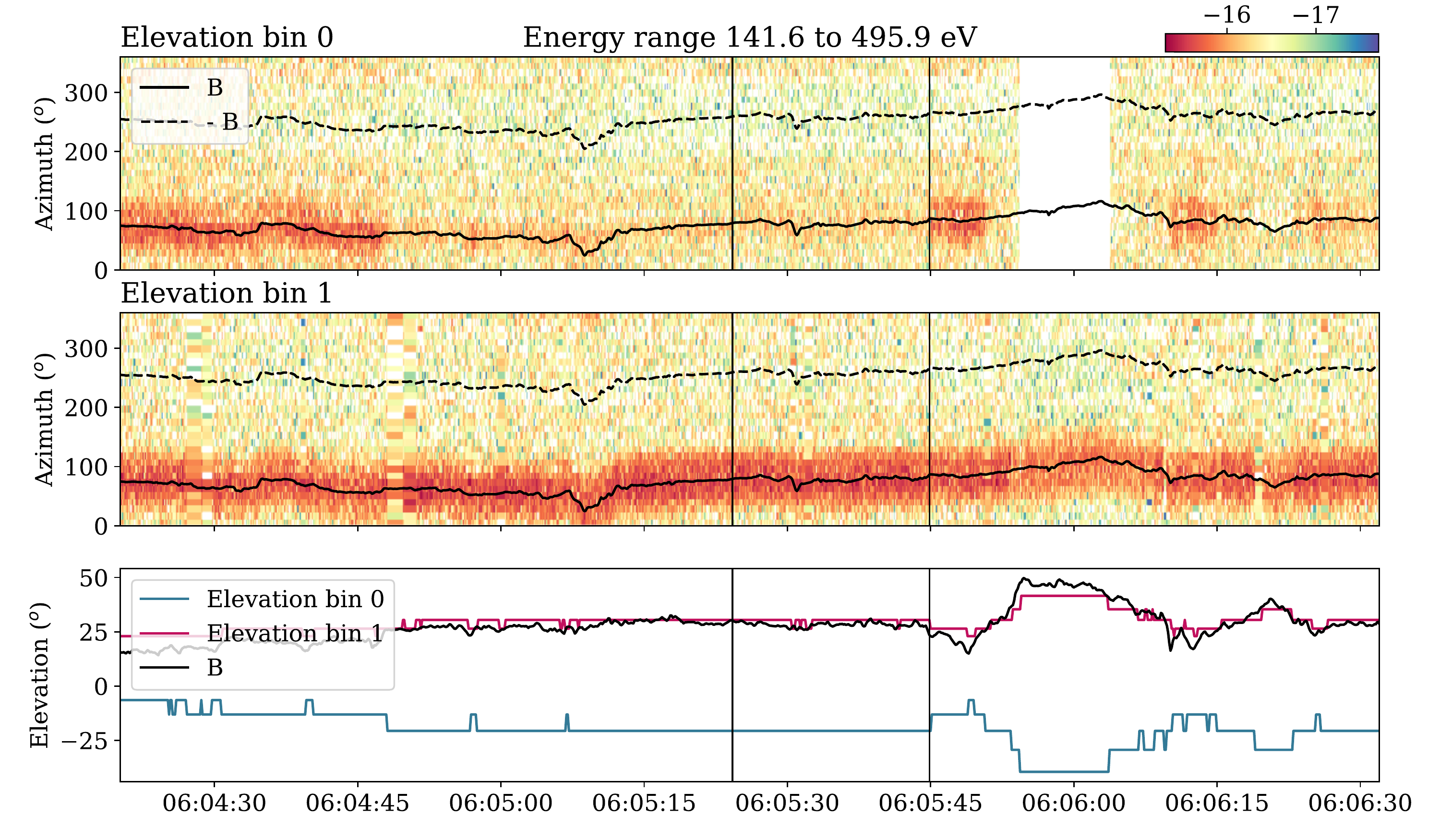}
    \caption{Time evolution of BM data in the instrument frame. The top two panels show the electron VDFs as functions of time and the EAS 1 azimuth angle for both of the selected elevation deflection states with a logarithmic colour scale. The bottom panel shows the sampled elevation angles. Black lines in all plots denote the direction of the magnetic field in the EAS 1 frame. }
    \label{fig:inst-frame}
\end{figure*}

Even though magnetic field measurements are calibrated on-board before being used by EAS, some differences occur between the on-board and the on-ground calibrated magnetic-field data. For our analysis, we use 8 Hz magnetic field measurements from MAG which have been recalibrated on-ground and are publicly available as L2 files on the SO archive\footnote{\url{http://soar.esac.esa.int/soar/##home}}. 

We present the case study based on one time interval of just over 2 min duration of BM data from June 24, 2020. We choose this time period because of the good quality of the EAS data -- high electron counts and good correlation between the on-board and on-ground magnetic fields -- and the presence of a clear signal of whistler waves in the associated RPW data.
The evolution of the electron VDF sampled in each of the two selected elevation bins is shown in the first two panels of Fig.~\ref{fig:inst-frame} as a function of the instrument azimuth angle and time. The third panel shows the corresponding elevation angles compared to the magnetic field direction. From the alignment between the magnetic-field direction and the automatically chosen elevation bin, we conclude that the magnetic field used by EAS on-board is close to the ground calibrated values. Strahl electrons, seen as a localised increase in electron flux, follow the magnetic field azimuth angle.

\begin{figure}
\centering
    \includegraphics[width=0.45\textwidth]{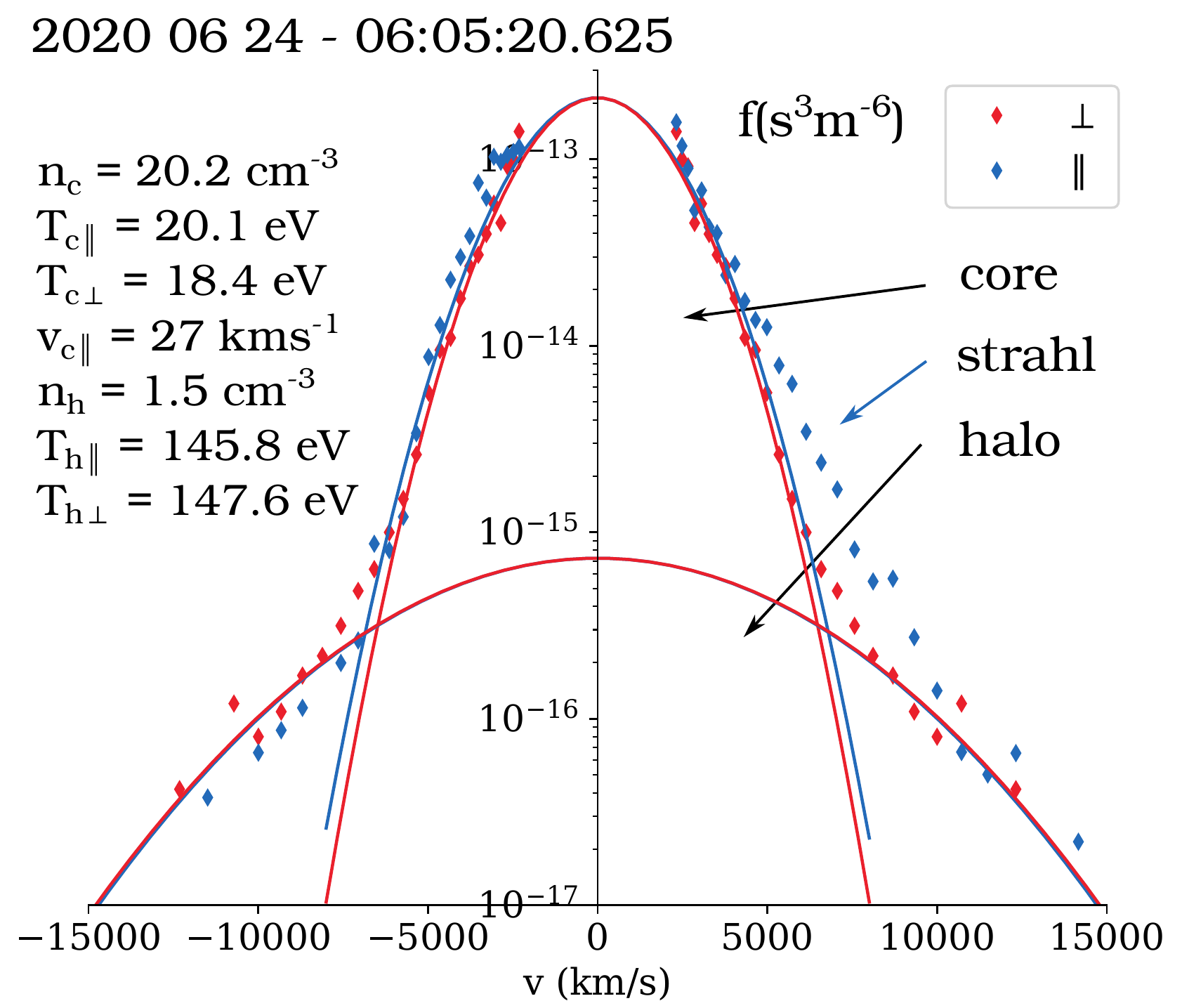}
    \caption{Example of an electron VDF integrated over two consecutive BM scans used for the core electron fit. Blue and red dots represent parallel and perpendicular cuts through the electron VDF, while the two curves show the core and the halo fits. The velocity is given in the instrument frame of reference and in the direction along the magnetic field.}
    \label{fig:vdf_fit}
\end{figure}

We fit our electron VDFs with a bi-Maxwellian function to obtain the electron core properties. Since the PAS sensor was not operating on the chosen day, we perform fits in the EAS magnetic-field aligned frame. We allow for a relative drift $v_ \mathrm{c\parallel}$ in our bi-Maxwellian distribution in the direction parallel to magnetic field. The bi-Maxwellian distribution is given by
\begin{equation}
    f_\mathrm c (v_{\perp}, v_\parallel)
= A_c \exp \left(- \frac{v_{\perp}}{w_{\mathrm c\perp}^2}
- \frac{(v_\parallel-v_{\mathrm c\parallel})^2}{w_{\mathrm c\parallel}^2} \right),
\label{eq:maxw}
\end{equation}
where  $A_\mathrm c$ is the normalisation factor, $w_\mathrm{c\perp}$ is the perpendicular core thermal velocity, $w_\mathrm{c\parallel}$ is the parallel core thermal velocity, and $v_\mathrm{c\parallel}$ is the core parallel drift velocity. These quantities are our fit parameters, from which we obtain the core parallel and perpendicular temperatures as
\begin{equation}
    T_\mathrm{c \parallel} = \frac{m_\mathrm e w_\mathrm{c\parallel}^2}{2k_\mathrm B}\;\;\;\text{and}\;\;\; T_\mathrm{c \perp} = \frac{m_\mathrm e w_\mathrm{c\perp}^2}{2k_\mathrm B}
\end{equation}
and the core density as
\begin{equation}
n_\mathrm c = A_\mathrm c \pi^{3/2} w_\mathrm{c\perp}^2 w_\mathrm{c\parallel}.
\end{equation}
In the equation above, $k_\mathrm B$ stands for the Boltzmann constant and $m_\mathrm e$ for the electron mass. 

To improve the signal-to-noise ratio, we perfom our fits on a moving window of two consecutive 2-dimensional gyrotropic BM VDFs. We use a least-square minimisation algorithm\footnote{scipy.optimize.leastsq (\url{https://docs.scipy.org/doc/scipy/reference/generated/scipy.optimize.leastsq.html})} provided by the Scipy Optimization package for Python \citep{virtanen}. Because the VDF values span over several orders of magnitude, we carry out our fits in logarithmic space ($\ln(f_\mathrm c)$). This technique decreases the large difference in the weight of the fitted data points. We use only energy bins between 15.3\,eV and 107.2\,eV to isolate the core population from the secondary electrons at lower energies and from the halo population at higher energies. We avoid the inclusion of strahl electrons  by excluding all data points within 30$^\circ$ pitch angle.

We normalise the core density obtained from the fit to the electron density obtained from the quasi-thermal-noise (QTN) technique derived from the plasma peak in the electric field power spectra \citep[and references therein]{MeyerVernet2017} from RPW \citep{Maksimovic2020}. This technique gives an accurate estimation of the electron density, which is limited in precision by discrete sampling frequency bins and time resolution. The average value of the total electron density during the presented time interval is 19 cm$^{-3}$ with an accuracy of 10\,\%. 

We fit the electron halo population using the same technique as described for the core. Due to the smaller signal-to-noise ratio in the halo energy range, we chose to fit the halo with a non-drifting Maxwellian, even though small drifts along the magnetic field direction in the plasma rest frame have been found in the past \citep{Stverak2009}. With one fitting parameter less than in the core case, the probability for a successful fit increases and the noise in the obtained halo temperatures ($T_\mathrm{h\perp}, T_\mathrm{h\parallel}$) decreases. We fit the halo to the difference between the observed VDF and the core fit, $f_\mathrm h = f - f_\mathrm c$, limited to the energy range between 162.8\,eV and  655.2\,eV. Measurements with pitch angles less than 30$^\circ$ are excluded to avoid the inclusion of the strahl electron population.

\subsection{Electromagnetic fluctuations}

Electromagnetic fluctuations are measured by the RPW triaxal coplanar electric antenna system (ANT), its biasing unit (BIAS), and a triaxal search-coil magnetometer (SCM) \citep{Jannet2021}. Their common, most exhaustive data product includes the recorded waveforms; however, due to their large size, full wave forms can only be downlinked for short periods of time. A complete overview of the wave activity at all times is assured by the spectral data product called Basic Parameters (BP), providing wave properties calculated on-board from time-averaged spectral matrices (ASM). In the present study, we only use the following wave parameters derived from the SCM measurements: the magnetic trace power spectrum, the degree of polarization, the wave ellipticity, and the wave normal vector. We also present the normalised electric-field power spectrum measured by RPW's electric-field antennas. \citet{Chust2021} provide a detailed description of the BP data products as well as a comparison of these reduced products with the full waveform data.

RPW also provides a snapshot waveform (SWF) data product during the selected period. SWF data consist of three times 2048 samples of magnetic and electric field fluctuations, available at different cadence (at best, every 22\,s, but here every 5\,min), acquired at three different sampling frequencies: 24\,576\,Hz, 4096\,Hz, and 256\,Hz. Whistler mode waves exhibit frequencies of order a few 10\,Hz, thus we investigate the 256\,Hz measurements consisting of an 8\,s-long waveform interval between 06:05:11.5 and 06:05:19.5. 

We identify frequency bands corresponding to the localised enhancements in the magnetic-field power spectra, which are characteristic for waves driven by kinetic instabilities. We perform a minimum-variance analysis of the bandpass-filtered data to obtain the wave normal vector $\hat{n}$. In our data intervals, the identified waves are circularly polarised, and their wavevector is almost aligned with the magnetic field (i.e., quasi-parallel propagation). The actual direction of propagation along $\hat{n}$ must be determined by considering the electric field measurements. \citet{Kretzschmar2021}  show that the overwhelming majority of whistler waves observed by SO propagate in the anti-sunward direction and exhibit  a "weak" phase deviation of 50$^\circ$ between the magnetic and electric field fluctuations that needs to be corrected. Applying this same correction here confirms the anti-sunward wave propagation.

Due to the lack of PAS data, we cannot directly transform the observed wave frequency from the SC frame to the plasma frame. The angle between the magnetic field vector and the radial direction is approximately 80$^\circ$ during the snapshot, which means that the magnetic field-aligned component of the quasi-radial solar-wind velocity is small, resulting in a small  Doppler shift of the frequency. We analyse both the electron data and the field data in the SC frame.

\section{Observation Results}
\label{sec:observations}
\subsection{Properties of the investigated time interval}
\begin{figure*}
\centering
    \includegraphics[width=0.93\textwidth]{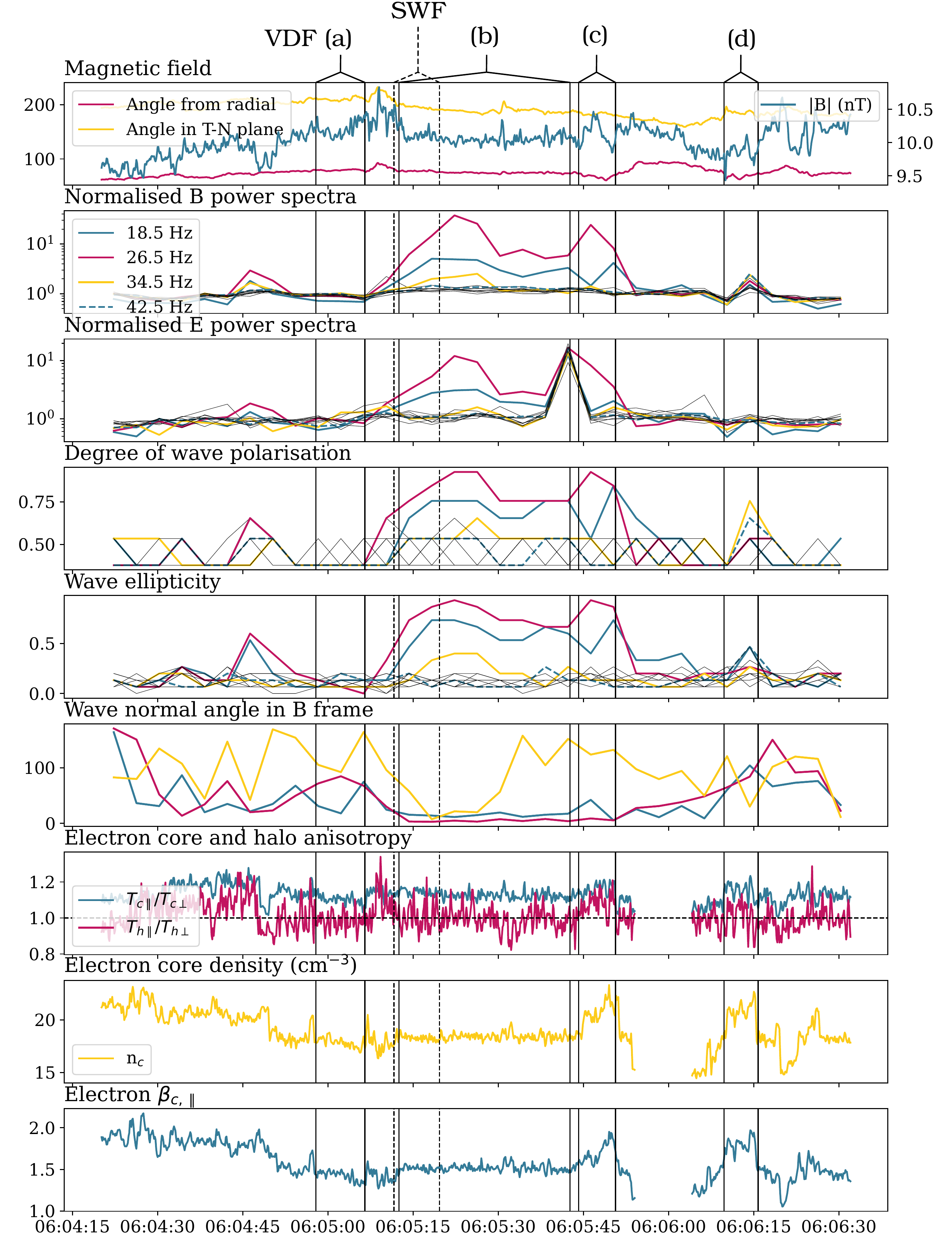}
    \caption{The plasma and EM field properties during the selected time interval. (1) Magnetic-field amplitude and direction obtained by MAG with a cadence of 8\,Hz; (2) Magnetic-field power spectra normalised with their median value during the selected interval. Each line denotes a separate frequency bin. We select the frequencies with enhanced fluctuations during the interval and colour them accordingly to the legend; (3) Electric-field power spectra presented in the same way; (4) Degree of polarisation, where 0 describes linear polarisation and 1 describes circular polarisation; (5) Wave ellipticity, where 0 describes a linear wave hodogram and 1 a cicular hodogram; (6) Direction of the wavevector with respect to the magnetic field; (7) Electron core temperature anisotropy ($T_\mathrm{c\parallel}/T_\mathrm{c\perp}$) and electron halo temperature anisotropy ($T_\mathrm{h\parallel}/T_\mathrm{h\perp}$); (8) Density;  (9) Electron core parallel beta. (2) to (6) are the product of on-board calculated basic parameters (BP) and have a time resolution of 4\,s \citep{Chust2021}.  }
    \label{fig:wave_params}
\end{figure*}

\begin{figure*}
    \includegraphics[width=1\textwidth]{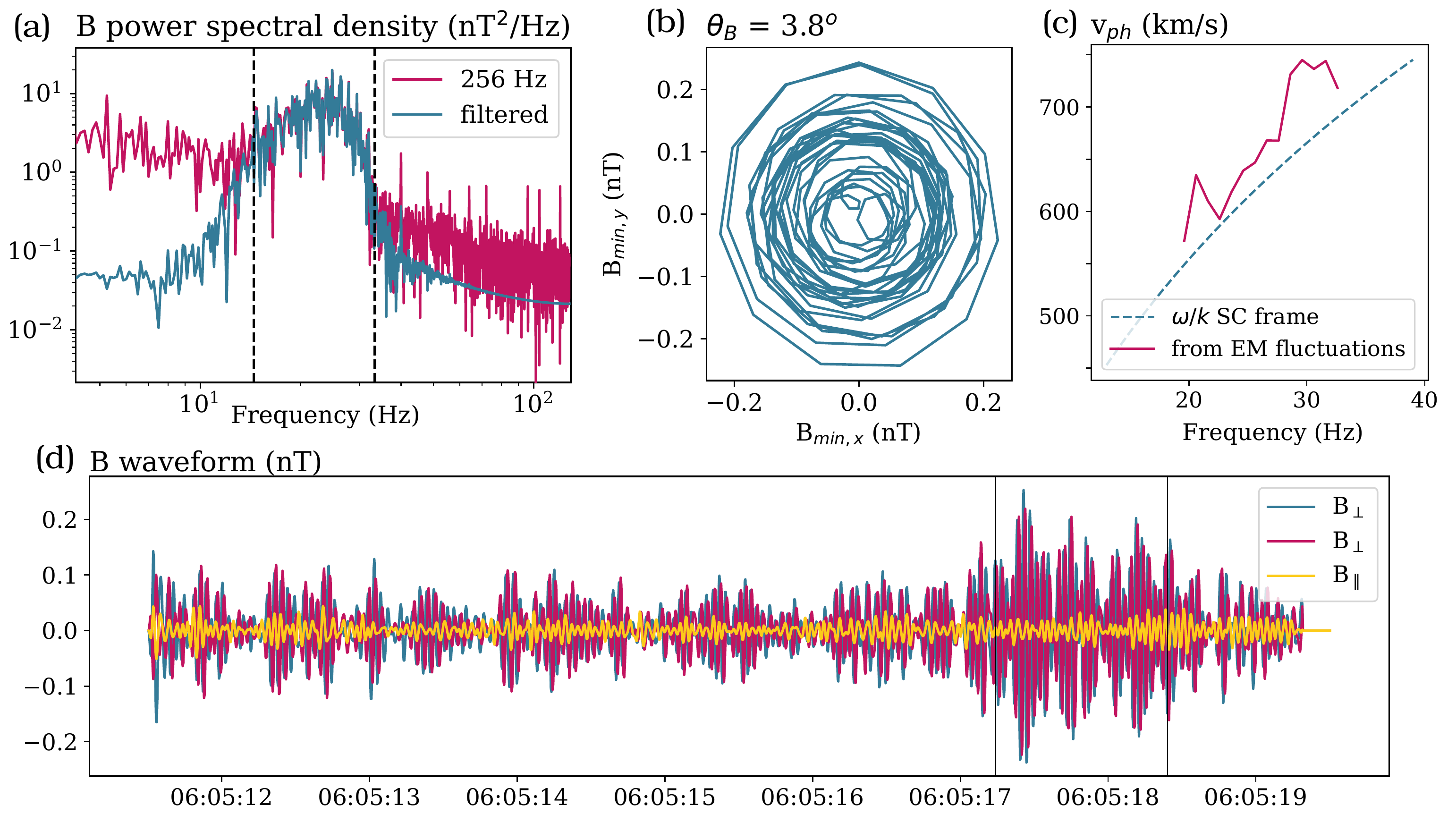}
    \caption{(a) Power spectral density of the 256\,Hz magnetic-field snapshot waveform from June 24th, 2020, 06:05:15.5 in pink and the band-pass filtered spectral density in blue; (b) A hodogram of the selected wave packet with the highest amplitude, representing the interval marked with black lines in the waveform data shown in (d); (c) Phase velocity determined from the electric and magnetic field fluctuations \citep{Kretzschmar2021} compared to the theoretical prediction ($\omega/k$) calculated from the wave frequency in the SC and solar-wind reference frame. We assume the solar wind with a radial velocity of 350\,km/s; (d) The snapshot waveform interval (SWF), where the $\vec B$ components are given in the background magnetic-field-aligned frame. The yellow colour denotes the parallel component. }
    \label{fig:swf}
\end{figure*}

We investigate a $\sim$ 2\,min long interval from June 24th, 2020, when SO was at a heliocentric distance of 112 $R_S$. Plasma and field properties during this interval are shown in Fig.~\ref{fig:wave_params}. Magnetic-field amplitude, and direction at 8\,Hz cadence shown in panel 1 stay approximately constant during the interval, while the fluctuations at higher frequencies vary with time. The normalised magnetic and electric field spectra in panels 2 and 3 show an increase in power at 18.5, 26.5 and 34.5\, Hz. Panels 4, 5, and 6 reveal that the increased fluctuations are almost circularly polarised  with a wavevector quasi-parallel to the magnetic field direction. Panel 7 displays the core electron temperature anisotropy and density, panel 8 the halo anisotropy, and panel 9 the electron core parallel beta, calculated as 
\begin{equation}
\beta_\mathrm{ec\parallel} = \frac{2 \mu_0 n_\mathrm c k_\mathrm B T_\mathrm{c \parallel}}{B^2},
\label{eq:ecpar}
\end{equation}
where $\mu_0$ is the vacuum permeability and $B$ is the magnetic field.

SWF data are available for the interval between the vertical black dashed lines in Fig.~\ref{fig:wave_params}. We show the analysis of the magnetic field components over the 8\,s snapshot in Fig.~\ref{fig:swf}. A large increase in fluctuation amplitudes occurs between 13 and 33\,Hz (in the SC frame) in the power spectral density (a). The fluctuations with bandwidth $\Delta f$ = 20\,Hz peak at $f_w$ = 24\,Hz, which corresponds to 0.085 $\omega_\mathrm{ce}$, where $\omega_\mathrm{ce}$ is the angular electron gyrofrequency. If we assume a typical radial solar-wind velocity of 350\,km/s, its projection into the direction of the magnetic field is 60\,km/s, resulting in a Doppler shift of 2.6\,Hz. In the waveform of the band-pass filtered $\vec B$ in the background magnetic field frame (d), the enhanced power of the fluctuations results from many separate wave packets. The wave amplitude $\mathrm{B}_w$ is approximately 0.15\,nT and the amplitude of the background field during the interval is $\mathrm B_0$ = 10.1\,nT. An example hodogram of the highest-amplitude wave packets (marked by black lines) is shown in (b). The average angle between the normal vector of the wave and the magnetic field direction is 3.8$^\circ$. Other wave packets show similar properties: strong alignment with the magnetic field  and right-hand circular polarisation.  

We determine the wave phase velocity, shown in pink in (c), from the magnetic and electric field measurements; assuming an effective antenna length of 14\,m in agreement with other studies \citep{Kretzschmar2021,Chust2021,steinvall2021}. Our results compare well with the theoretical expectation of $\omega / k$ plotted in blue, which is discussed further in the following section. 

\begin{figure*}
    \includegraphics[width=1\textwidth]{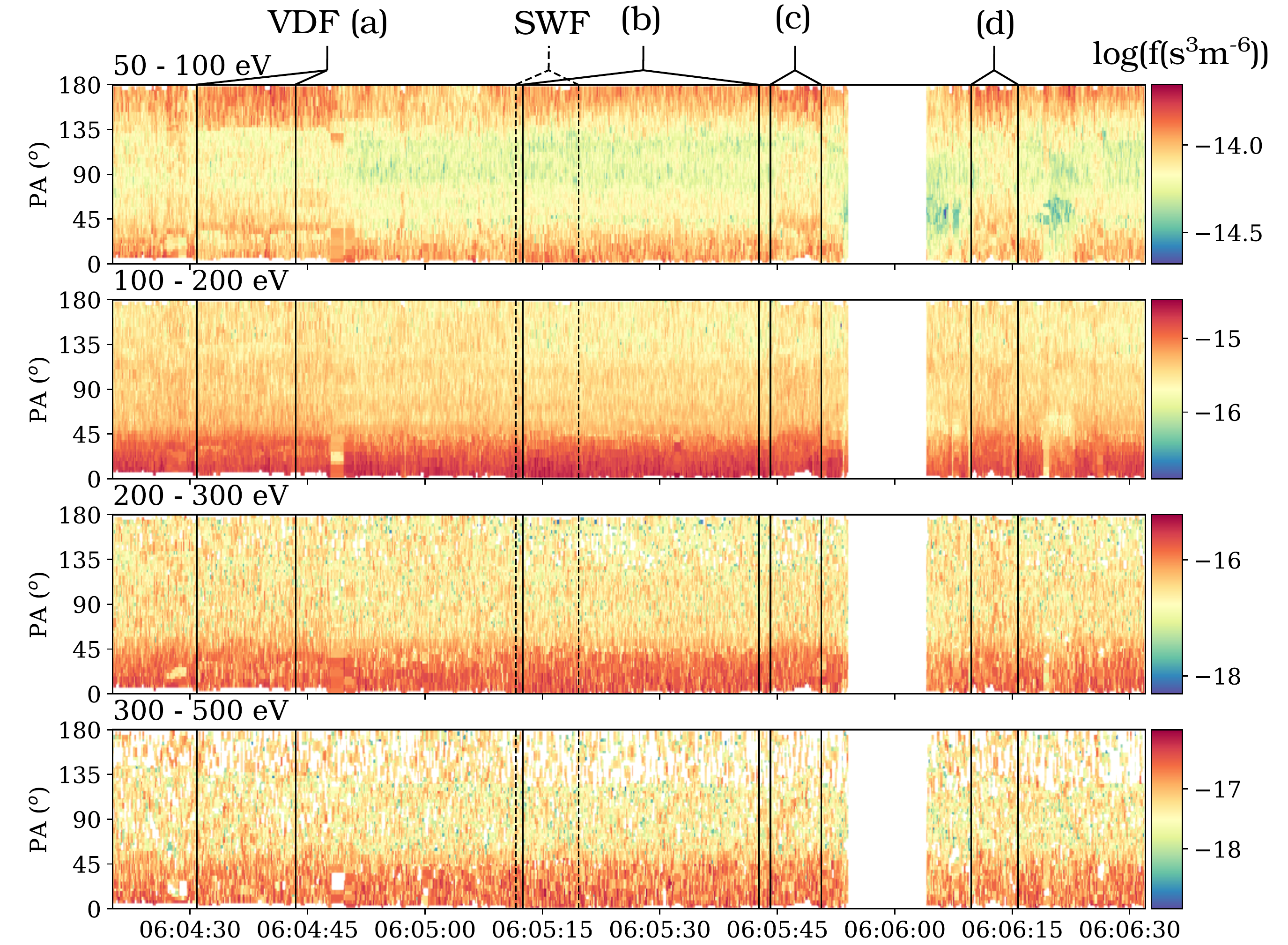}
    \caption{BM pitch-angle distributions (PADs) averaged over different energy ranges (indicated in the title of each plot). The colour coding represents the logarithm of the electron VDF.  The vertical black lines are the same as in Fig \ref{fig:wave_params} and denote the intervals of integration of electron VDFs shown in Fig.~\ref{fig:vdfs} and the snapshot waveform interval. The gap in the data set beginning at around 06:06:00 is due to the extreme elevation setting in the relevant EAS head (see Fig.~\ref{fig:inst-frame}).}
    \label{fig:pads}
\end{figure*}

\begin{figure*}
\centering
    \includegraphics[width=0.9\textwidth]{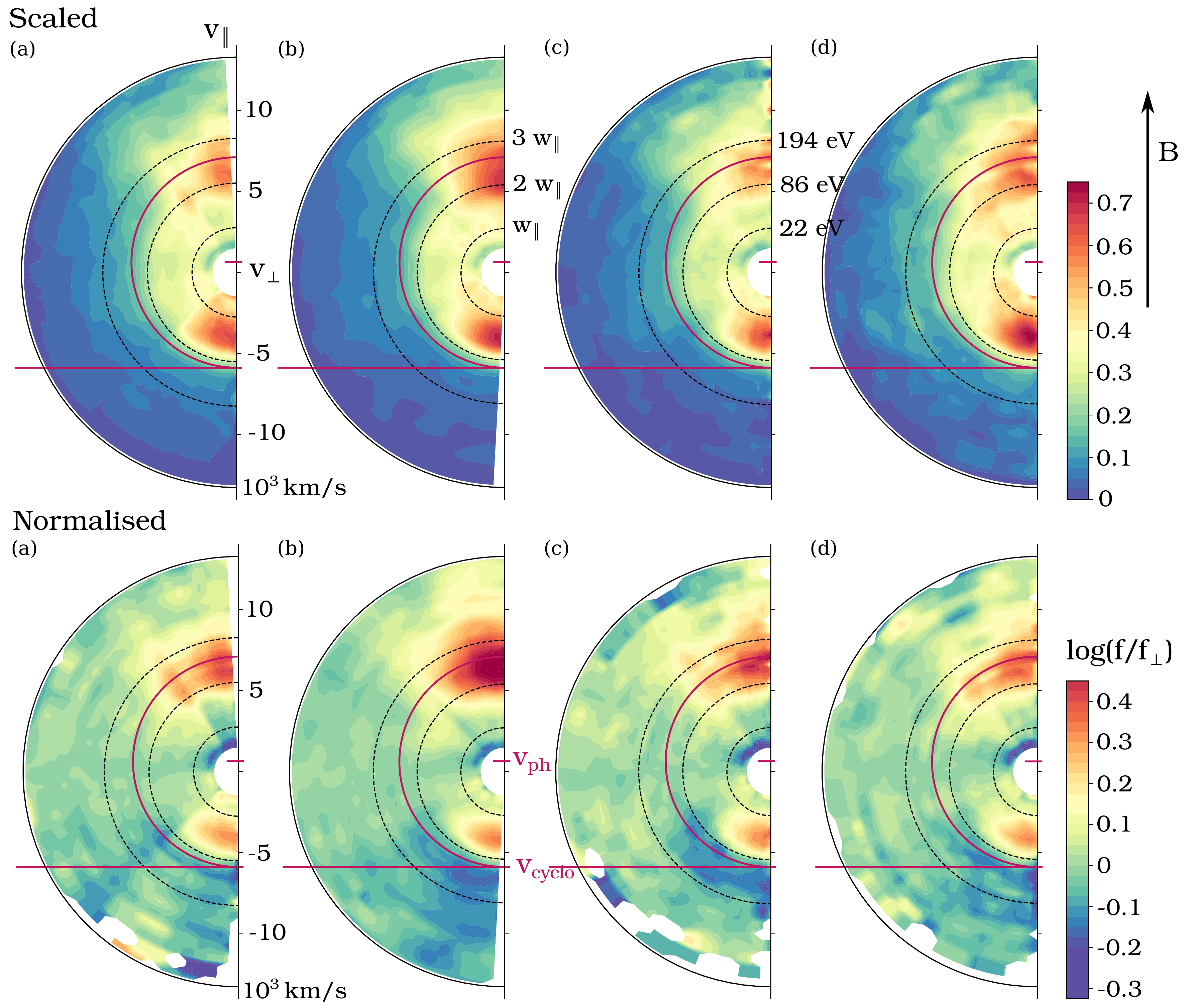}
    \caption{BM electron VDFs averaged over the time periods corresponding to those shown in Figs.~\ref{fig:wave_params} and \ref{fig:pads}, in two representations highlighting departures from an isotropic distribution function. The first row displays the \emph{scaled} VDFs, where the values in each energy belt are scaled between 0 and 1. The second row displays the \emph{normalised} VDFs, where all  values are normalised to the cut along the perpendicular direction. The magnetic-field direction is indicated with an arrow on the right-hand side and is aligned with the Y-axis. Black dashed semi-circles denote 1, 2, and 3 $w_\mathrm c$. }
    \label{fig:vdfs}
\end{figure*}

We show the time evolution of the BM electron VDF through pitch-angle distributions (PADs) averaged over different electron energies in Fig.~\ref{fig:pads}. In the lowest energy range, 50 - 100\,eV, we observe two features in the parallel and anti-parallel directions with respect to the magnetic field, while for higher energies only one of the features -- the strahl electrons -- remains. In order to compensate for the noise in single electron-VDF measurements, we integrate the VDFs over selected time periods during which the PADs are similar. We show these integrated VDFs as functions of $v_\parallel$ and $v_\perp$ in Fig.~\ref{fig:vdfs} through a \emph{scaled} and \emph{normalised} representation, highlighting the gyrotropic non-isotropic features \citep{Behar2020}. In \emph{scaled} VDFs, presented in the first row, each energy bin -- each circular belt in $(v_\parallel, v_\perp)$ parameter space -- is scaled to a value between 0 and 1, where 1 corresponds to the maximum value of the VDF in the given energy bin. With this representation, we remove the information about the absolute value of the VDF and its strong gradient with energy. The benefit of this representation is the exposure of smaller anisotropic features at all energies. In cases in which two features arise in the same energy bin, the scaled VDFs can be misleading as they focus on the bigger feature. The second row presents \emph{normalised} VDFs, obtained by normalising the VDF with the cut along the perpendicular direction $f(v_\perp,v_\parallel=0)$. Pitch-angle directions in which the distribution function is less than $f(v_\perp,v_\parallel=0)$, appear in blue, and those in which the distribution function is greater than $f(v_\perp,v_\parallel=0)$ appear in red. 

We show the VDFs in the instrument frame and thus expect a small drift of the electron core in the direction opposite to the heat flux. This drift is visible in Fig.~\ref{fig:vdfs} as a slight depletion in the positive $v_\parallel$ direction, and a slight overdensity in the negative $v_\parallel$ direction in the thermal electron energy range, within the circle marking 1 $w_\mathrm{c\parallel}$. In the velocity range between 1 and 2 $w_\mathrm{c\parallel}$, we observe an overdensity in the sunward / anti-strahl direction. This feature also exists in many electron VDFs measured by PSP \citep[e.g., in Fig.~1 by][]{Halekas2020}. Just below 2 $w_\mathrm{c\parallel}$, we see a transition between the overdensity and the suprathermal deficit in the sunward direction. In the anti-sunward direction, we detect the beginning of the strahl component at similar energies. Most of the strahl electrons have velocities between 2 and 3 $w_\mathrm{c\parallel}$. 

Example (a) represents the VDF during the time period before the wave detection, where $n_\mathrm c$ and $T_\mathrm{c\parallel}/T_\mathrm{c\perp}$  are slightly greater than during the rest of the interval, resulting in higher $\beta_\mathrm{ec}$. We also observe a subtle increase in electron $T_\mathrm{h\parallel}/T_\mathrm{h\perp}$. Above 3 $w_\mathrm{c\parallel}$, the strahl shows signs of scattering towards larger pitch-angles, and the halo populates all pitch angles. The sunward deficit is not pronounced. Example (b) represents the electron VDF during the first part of the whistler-wave period, for which SWF data are available. During this interval, $n_\mathrm c$, $T_\mathrm{c\parallel}/T_\mathrm{c\perp}$ and $T_\mathrm{h\parallel}/T_\mathrm{h\perp}$ decrease, and $\beta_\mathrm{ec}$ drops to 1.5. The VDF exhibits a denser, clearly defined strahl, persisting somewhat above 3 $w_{c\parallel}$. The sunward deficit is more pronounced. Example (c) represents the electron VDF during the interval in which the waves are still present, but an increase is observed in $T_\mathrm{c\parallel}/T_\mathrm{c\perp}$, $T_\mathrm{h\parallel}/T_\mathrm{h\perp}$, $n_\mathrm c$ and $\beta_\mathrm{ec}$. This VDF exhibits a weaker strahl and the halo electrons are present at all pitch angles above 3 $w_\mathrm{c\parallel}$. However, between 1 and 2 $w_\mathrm{c\parallel}$ we still observe the sunward deficit. Example (d) describes an interval during which enhanced field fluctuations appear at higher frequencies; however, these fluctuations are short lasting, described by only one point of the BP data. Thus, we do not discuss them in more detail. The electron VDFs in this case are similar to those in (c), except that the deficit in (d) is less pronounced and extends to higher energies.

\subsection{The quasi-parallel whistler instability driven by the sunward deficit}

Simultaneous observations of waves and high-cadence electron VDFs from SO allow us to test our prediction for the instability scenario of quasi-parallel whistler waves driven by the sunward electron deficit in solar-wind data. 

From the measured wave frequency, $\omega$, we estimate the associated wavenumber  $k$ based on the cold-plasma dispersion relation for parallel-propagating whistler waves
\begin{equation}
\label{eq:dispersion}
    k = \frac{\omega}{c}\sqrt{1 - \frac{ \omega_\mathrm{pe}^2 / \omega^2}{1 - \omega_\mathrm{ce}/\omega}},
\end{equation}
where
\begin{equation}
\omega_\mathrm{pe} = \sqrt{\frac{n_\mathrm c  e^2}{m_\mathrm e \epsilon_0}}
\end{equation}
is the electron plasma frequency.
For this calculation, we use the peak frequency determined from the SWF data in the SC frame. We obtain a wave phase velocity of $v_\mathrm{ph} \sim 608 km/s$, and thus a resonant speed of $v_\mathrm{cyclo} \sim -6490 km/s$ according to Eq.~(\ref{eq:res}).

The wave properties obtained from the SWF interval allow us to evaluate whether the interaction between the observed waves and solar wind electrons  is compatible with the assumptions of the quasilinear theory. The quasilinear theory is applicable, if the width of the wave spectrum is sufficiently large \citep{SagdeevGaleev1969, Tong2019stat}:

\begin{equation}
    \frac{\Delta f}{f_w} \gg \left( \frac{\mathrm B_w}{\mathrm B_0} \right) ^{1/2} \left (\beta_\mathrm{ec\parallel}\frac{\omega/\omega_\mathrm{ce}}{1-\omega/\omega_\mathrm{ce}} \right) ^{1/4}.
    \label{eq:ql}
\end{equation}

Using the wave parameters presented in Section \ref{sec:observations} we obtain 0.83 for the left-hand side of Eq.~(\ref{eq:ql}) and 0.074 for the right-hand side of Eq.~(\ref{eq:ql}), which confirms the applicability of the quasilinear approach.

In order to test our scenario outlined in Section~\ref{sect:prediction}, we overplot $v_\mathrm{cyclo}$ in our phase-space plots in Fig.~\ref{fig:vdfs} as horizontal pink lines. The resonance speed coincides well with the sunward electron deficit, suggesting that electrons associated with the sunward deficit can indeed fulfil the cyclotron-resonance condition with quasi-parallel whistler waves. In fact, the deficit is more pronounced during the intervals with increased amplitudes of quasi-parallel whistler waves (examples (b) and (c) in bottom row of Fig.~\ref{fig:vdfs}), yet not present in the example (a) sampled in absence of whistler-wave activity. 

\begin{figure}
\centering
    \includegraphics[width=0.5\textwidth]{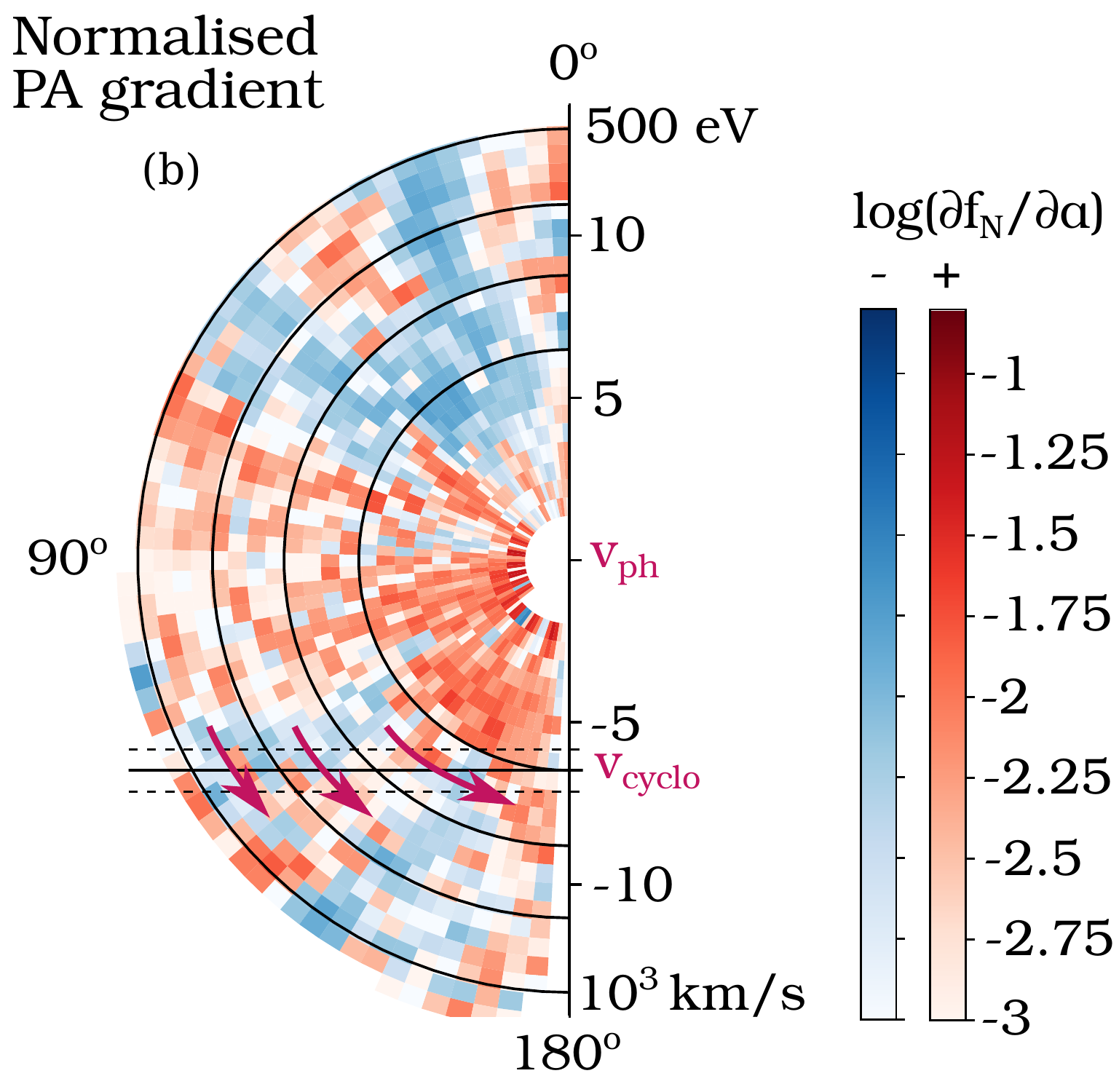}
    \caption{Normalised pitch-angle gradient of the electron VDF from Fig.~\ref{fig:vdfs} (b) in the frame centred on the phase speed of the quasi-parallel whistler waves. Positive pitch-angle gradients $\partial f_N/\partial \alpha^\prime$ are shown in red and negative pitch-angle gradients in blue. They are calculated separately for each velocity bin from the normalised VDF with respect to the pitch-angle increasing from the +Y-axis (0$^\circ$), to the -Y-axis (180$^\circ$). The black semi-circles show the constant energy curves in the wave frame, and the horizontal black lines mark $v_{\parallel}=v_\mathrm{cyclo} \pm 10\%$. Pink arrows indicate the electron diffusion paths according to the scenario described in Section~\ref{sect:prediction}.  }
    \label{fig:gradient}
\end{figure}

This first comparison does not yet reveal which of the two scenarios shown in Fig.~\ref{fig:schematics} applies. To determine the direction of the quasilinear electron diffusion in velocity space, we calculate the pitch-angle gradient in the wave rest frame, which is presented in Fig.~\ref{fig:gradient} for example (b). We obtain the gradient by first shifting the electron VDF to the wave frame, centred on $v_\mathrm{ph}$. For each of the velocity bins, we then calculate $\partial f_N / \partial \alpha^\prime$, where $\alpha^\prime$ is the pitch-angle, starting with 0$^\circ$ in the strahl direction, increasing towards the deficit, and $f_N$ is the normalised VDF. The strahl region appears as a negative gradient (blue) because the phase-space density decreases between 0$^\circ$ and 90$^\circ$. We also find negative pitch-angle gradients around $v_\mathrm{cyclo}$ marked with a horizontal black line. This finding indicates that resonant electrons in this region of velocity space indeed diffuse from larger to smaller $v_\perp$, in the direction marked with pink arrows. Fig.~\ref{fig:gradient} thus indicates that the observed mechanism corresponds to the case shown in Fig.~\ref{fig:schematics} (b), in which the resonant electrons lose kinetic energy and thus drive the whistler waves unstable.

We only find a clear correspondence between the observations and our scenario for example (b), as the normalised gradient is too noisy in the other VDFs to make conclusions about the quasilinear electron diffusion. In our future studies, we intend to improve the data analysis technique to obtain smooth pitch-angle gradient distributions, which are an important tool for the stability analysis of VDFs. Better resolution can result from using a higher-order interpolation technique instead of the nearest-neighbour interpolation used in the present work \citep{Behar2020}. The accuracy will also improve in the electron VDFs measured during upcoming SO orbits, as EAS settings are being adapted to reach its optimal performance.

Following the proposed instability scenario, electrons diffuse towards the suptrathemal deficit in the velocity space and tend toward filling it up. \citet{Halekas2020} model the electron heat flux in the solar wind with three contributions: the core drift leading to  sunward heat flux, while the strahl and the sunward deficit represent antisunward contributions, which are often larger in amplitude than the sunward contribution. Therefore, the proposed instability, regulating the electron VDF by filling the sunward deficit, reduces the total electron heat flux, potentially to a significant degree. A quantification of its impact is beyond the scope of this work.

\section{Discussion and Conclusions}
\label{sec:discussion}

We propose an instability scenario in which quasi-parallel whistler waves are created self-consistently with the quasilinear diffusion of resonant electrons associated with the sunward electron deficit in phase-space. The diffusion is made possible by a non-Maxwellian deviation of the pitch-angle gradients in the supratermal deficit, which has been recently observed in the anti-strahl direction in near-Sun solar wind.

We outline a theoretical prediction for the resonance condition of quasi-parallel whistler waves with electrons in the sunward electron deficit. We find that, if the sunward deficit is strong enough, the electron VDFs in the near-Sun solar wind can fulfil all conditions for a resonant instability of quasi-parallel whistler waves.
We test our prediction based on simultaneous observations of high-cadence electron VDFs and quasi-parallel whistler waves from SO. We find that the electron velocity corresponding to the cyclotron resonance with the observed waves coincides with the velocity-space region associated with the electron deficit. In the same region of phase-space, we find negative pitch-angle gradients in the wave rest frame, which is consistent with the direction of the quasilinear diffusion of electrons in our instability scenario.

We discuss the possible mechanisms responsible for the creation of the deficit. The first (and to us most probable) explanation is that the deficit is a consequence of the weakly-collisional radial expansion of the solar wind \citep{Halekas2021, Bercic2021ambip}. In this scenario, the deficit is a remnant of the collisionless exospheric electron cutoff \citep{Jockers1970, Lemaire1970, LemaireJosephandScherer1971} smoothened by  Coulomb collisions. Electron VDFs obtained in  kinetic solar-wind models that account for Coulomb collisions predict a sunward deficit similar to the one observed in the near-Sun solar wind \citep{pierrard2001, Landi2012b, Landi2014b, Bercic2021}.

The second possibility is that the deficit results from the scattering of the strahl population beyond a pitch angle of 90$^\circ$. This behaviour is  observed in a numerical study of strahl scattering  \citep{Micera2020}. The simulation starts with a VDF consisting of only a core and a strahl population. The strahl first triggers the oblique whistler instability (O-WHFI), which results in the self-induced scattering of the strahl electrons towards larger pitch-angles \citep{Verscharen2019scat, Vasko2019}. These electrons are then scattered  to 90$^\circ$ pitch-angle, increasing the phase-space density around $v_\parallel = 0$. At this point, the pitch-angle gradient for $\alpha^\prime > 90^\circ$ becomes negative and allows for electrons to diffuse from larger to smaller $v_\perp$, although the details of the resonant mechanism at this point are still unclear. Self-induced scattering of the strahl electrons can only create the deficit in the halo energy range, and cannot explain the deficit in the core population. After the saturation of the O-WHFI, a secondary instability is triggered in the simulation preformed by \citet{Micera2020}, resulting in quasi-parallel whistler waves propagating along the strahl direction. This second part of the \citet{Micera2020} scenario aligns well with our results and thus describes an alternative pathway leading to the conditions required for our instability mechanism.

The third possibility is that the deficit is created by the mechanism described in Fig.~\ref{fig:schematics} (a) itself. This scenario corresponds to the results of a nonlinear evolution of the WHFI as shown with a particle-in-cell simulation by \citet{Kuzichev2019}. In this scenario, the WHFI itself is driven by relative drift between the core and the halo populations and generates quasi-parallel whistler waves propagating in the direction of the heat flux \citep{Gary1975b, Gary1994,Lacombe2014,Kajdic2016a, Tong2019, Jagarlamudi2021}. These waves then deform the electron VDF through resonant damping forming the sunward deficit. However, the deficit found by \citet{Kuzichev2019} is small, and does not significantly change the total heat flux of the overall electron distribution. This result differs from the observations, which show that the sunward deficit can be responsible for more than a third of the anti-sunward heat flux contribution \citep{Halekas2020}. Albeit worthwhile for a complete understanding of the relevant processes, we leave a further investigation of these scenarios for a future study. We note, however, that our results and the presence of quasi-parallel whistler waves driven by the sunward electron deficit are consistent with all of these scenarios.

We conclude that the instability driven by the sunward deficit can create the observed quasi-parallel whistler waves and lowers the total heat flux stored in the electron VDF.

\begin{acknowledgements}
      Solar Orbiter is a space mission of international collaboration between ESA and NASA, operated by ESA. Solar Orbiter Solar Wind Analyser (SWA) data are derived from scientific sensors which have been designed and created, and are operated under funding provided in numerous contracts from the UK Space Agency (UKSA), the UK Science and Technology Facilities Council (STFC), the Agenzia Spaziale Italiana (ASI), the Centre National d’Etudes Spatiales (CNES, France), the Centre National de la Recherche Scientifique (CNRS, France), the Czech contribution to the ESA PRODEX programme and NASA. Solar Orbiter SWA operations work at UCL/MSSL is currently funded under STFC grants ST/T001356/1. L.~B., C.~J.~O., and D.~V. are supported by STFC Consolidated Grant ST/S000240/1. D.~V.~is supported by STFC Ernest Rutherford Fellowship ST/P003826/1. R. T. Wicks is funded by STFC grant ST/V006320/1. The RPW instrument has been designed and funded by CNES, CNRS, the Paris Observatory, The Swedish National Space Agency, ESA-PRODEX and all the participating institutes.
\end{acknowledgements}

\bibliographystyle{bibtex/aa}
\bibliography{bibliography.bib}

\end{document}